\documentclass[12pt,a4paper]{iopart}
\usepackage{graphicx}
\usepackage{amssymb}
\usepackage{amsthm}
\usepackage{hyperref}
\usepackage{color}
\usepackage{xcolor}
\usepackage{lineno}
\modulolinenumbers[1]

\begin{document}

\title[Crystal defects formation in Mo]{Computational study of crystal defects formation in Mo by machine learned molecular dynamics simulations}


\author{F. J. Dom\'inguez-Guti\'errez$^1$\footnote[4]{Present address: NOMATEN Centre of Excellence, National Center for Nuclear Research, ul. A. Sołtana 7, 05-400 Swierk/Otwock, Poland}, J. Byggm\"astar$^2$, K. Nordlund$^2$, F. Djurabekova$^{2,3}$, 
and U. von Toussaint$^1$}

\address{$^1$ Max-Planck Institute for Plasma Physics, Boltzmannstrasse 2, 85748
 Garching, Germany.}
 \address{$^2$ Department of Physics, University of Helsinki, Helsinki, PO Box 43, FIN-00014, Finland.}
 \address{$^3$ Helsinki Institute of Physics, Helsinki, Finland.}
\ead{javier.dominguez@ipp.mpg.de}

\begin{abstract}
In this work, we study the damage in crystalline molybdenum material 
samples due to neutron bombardment in a primary knock-on atom range of 
0.5-10 keV at room temperature.
We perform machine learned molecular dynamics (MD) simulations 
with a previously developed interatomic potential based on the Gaussian Approximation 
Potential (GAP) framework.
We utilize a recently developed software workflow for fingerprinting and 
visualizing defects in damage crystal structures to analyze the damaged 
Mo samples by computing the formation of point defects during and after 
a collision cascade.
As a benchmark, we report results for the total number of Frenkel 
pairs (a self-interstitial atom and a single vacancy) formed and atom 
displacement as a function of the PKA energy. 
A comparison to results obtained by using an Embedded
 Atom Method (EAM) potential is presented to discuss the advantages and 
limits of the machine learned MD simulations. 
The formation of Frenkel pairs follows a sublinear scaling law related to the PKA energy 
with $E^{0.54}_\mathrm{PKA}$ to the GAP MD results and $E^{0.667}_\mathrm{PKA}$ for the EAM 
simulations.
Although the average number total defects is similar for both methods, we notice that
MD potentials model different atomic geometries for the complex point defects, where the 
formation of crowdions is more favorable for the GAP potential. 
Finally, ion beam mixing results for GAP MD simulations are reported 
and discussed.

\end{abstract}
\noindent{\it Keywords \/}: Molybdenum, MD simulations, descriptor vectors, machine learning, material damage analysis, Gaussian Approximation potentials \\
\submitto{MSMSE}

\maketitle
\section{Introduction}
\label{intro}

The design of next generation of fusion machines needs experimental 
exploration of different plasma facing materials (PFM) candidates and 
the support and validation of numerical modeling \cite{doi:10.1063/1.2336465,wirth_hu_kohnert_xu_2015}. 
Molybdenum has been selected as a candidate for a PFM due its high 
melting points, good resistance to deformation, and 
low sputtering yield under plasma irradiation \cite{Eren_2011}.
Mo is used for diagnostic mirrors in fusion machines to deal with 
the harsh plasma environment and fluxes of neutrals and neutron radiation \cite{doi:10.1063/1.2336465,LITNOVSKY20071395}. 
In order to guide PFM experiments and to understand the mechanism of the 
interaction of plasma with materials, atomistic simulations based on the 
molecular dynamics (MD) method can be performed \cite{Nor17b,Nor18,BOLT200243}. 
Serving, at the same time, to save laboratories and financial resources for carrying out 
the PFM experiments.

Several transition metals and alloys are traditionally modeled by the 
embedded atom method (EAM) potentials in MD simulations 
\cite{PhysRevB.29.6443,Nor17b,AcklandGJ,Salonen_2003}, reproducing many 
material properties in good agreement with those measured experimentally. 
However, EAM and other traditional potentials are limited to fixed functional forms \cite{PhysRevB.29.6443,doi:10.1063/1.2336465} and can wrongly model some point
defects that are energetically unstable, or lack physical meaning in material
damaging processes \cite{DOMINGUEZGUTIERREZ2020100724}.
For this reason, interatomic potentials computed by using machine learning (ML) methods 
are now increasingly used to perform MD simulations with an accuracy close to 
Density Functional Theory \cite{Jesper_GAP,PhysRevLett.104.136403,doi:10.1080/21663831.2020.1771451}.
Being systematically improved towards the accuracy of the training data set.
The goal of the present work is to numerically model the damage in crystalline Mo samples 
due to plasma irradiation by utilizing the recently developed ML interatomic potentials 
by Byggm\"astar et al. \cite{2006.14365}. 
Thus, we perform MD simulations to emulate neutron bombardment at intermediate primary 
knock-on atom energies providing an understanding about the modeling of the re-crystallization
process after the collision cascade, which has been an issue for numerical simulations based 
on fixed functional forms \cite{DOMINGUEZGUTIERREZ2020100724,Jesper_GAP}.

Our paper is organized as follows: in Sec. \ref{methods} 
we briefly discuss the theory to develop the machine learned (ML) potential 
\cite{2006.14365} for Mo, and the software workflow for fingerprinting and visualizing
defects in damaged crystal structures (FaVAD) \cite{2004.08184,favad} that is applied
to quantify and classify the damage in Mo samples \cite{Jav_UvT}.
Our results for the total number of points defects, Frenkel pairs and atomic displacement 
are presented in Sec. \ref{results}. 
We examine the limitations and advantages of our new ML interatomic potential by 
comparing to MD simulations results obtained by EAM potentials. 
Finally, in section \ref{conclusions}, we provide concluding remarks.

\section{Methods}
\label{methods}

Interatomic potentials based on machine learning (ML) methods are
not restricted to an analytical form and can be systematically improved
towards the accuracy of the training data set. 
In order to model collision cascades, the ML potential must be able 
to treat realistic short-range dynamics defined by its repulsive part. 
In addition, the correct structure of the liquid phase and 
re-crystallization process should be well 
described, to accurately emulate atomic mixing together with defect
creation and annihilation during the collision cascade.
In this work, we use the ML interatomic potential for Molybdenum that was
recently developed \cite{2006.14365} within the Gaussian Approximation Potential (GAP)
framework \cite{PhysRevLett.104.136403,PhysRevB.87.184115}. 
Here, the total energy of a system of $N$ atoms is expressed as 
\begin{equation}
    E_{\textrm{tot}} = \sum^N_{i<j} V_{\textrm{pair}}(r_{ij}) + 
    \sum^{N_d}_{i} E^i_{\textrm{GAP}},
\end{equation}
where $V_\mathrm{pair}$ is a purely repulsive screened Coulomb
potential, and $ E_{\textrm{GAP}}$ is the machine learning
contribution. $ E_{\textrm{GAP}}$ is constructed using a 
two-body and the many-body Smooth Overlap of Atomic Positions 
(SOAP) descriptor \cite{PhysRevLett.104.136403}. 
$N_d$ is the number of descriptor environments for the $N$-atom system 
(i.e. number of pairs for the two-body descriptor and number of atoms for
the many-body descriptor). 
The ML part of the potential is given by

\begin{eqnarray}
E^i_{\textrm{GAP}}  & =  \delta_\textrm{2b}^2 \sum_j^{M_{\textrm{2b}}} 
\alpha_{j,\textrm{2b}} K_{\textrm{2b}} (\vec q_{i,\textrm{2b}}, 
\vec q_{j,\textrm{2b}}) \nonumber \\
 & + \delta_\textrm{mb}^2 \sum_j^{M_{\textrm{mb}}} \alpha_{j,\textrm{mb}}
         K_{\textrm{mb}} (\vec q_{i,\textrm{mb}}, \vec q_{j,\textrm{mb}}),
\end{eqnarray}
where $\delta^2_{\textrm{2,mb}}$ are prefactors that set the energy ranges of the ML predictions; $K_{\textrm{2,mb}}$ is the kernel function representing the similarity 
between the atomic environment of the $i$-th and $j$-th atoms;
$\alpha$ is a coefficient obtained from the fitting process; 
and $\vec q$ is the normalized descriptor vector of the local 
atomic environment of the $i$-th atom (See Sec 2.3).
In the computation of the ML potential the descriptors for two bodies, 2b, 
is utilized to take into account most of the interatomic bond energies, 
while the many-body, mb, contributions are treated by the SOAP descriptor.
More details about the computation of the ML potentials for Mo can be found 
in Ref. \cite{2006.14365}.

\subsection{MD simulations}
\label{sec:md_simulations}

Machine learning based MD simulations are performed to model neutron bombardment
processes at different PKA energy values to analyze damage in 
crystalline materials. 
We compare results to those obtained by the Embedded Atom Method (EAM)  
\cite{Salonen_2003} by considering similar numerical parameters to explore 
the advantages and limitations of the new GAP interatomic potential.
For this, we first define a simulation box as a pristine Mo crystalline 
sample based on a body-centered-cubic (bcc) unit cell with a lattice 
constant of $a = 3.16$ \AA{} according to DFT calculations for computing the
GAP potentials and those reported in the literature
\cite{PhysRevB.85.214121}.
This value is slightly higher than the experimental measurement
\cite{WBPearson}. 
Then, the numerical sample is prepared by a process of energy optimization
and thermalization to 300 K using the Langevin thermostat, with the time 
constant of 100 fs. \cite{DOMINGUEZGUTIERREZ201756}. 
The room temperature is used in our work to perform numerical simulations as 
close as possible to the experiments of material damaging
\cite{doi:10.1063/1.2336465}.
The MD simulation is started by assigning a kinetic energy to a Mo atom located 
at the center of the numerical sample in a range of 0.5-10 keV of PKA.
For each PKA energy value, the projectile travels on ten different crystal 
orientation: 
$\langle 0 0 1 \rangle$, $\langle 1 1 0 \rangle$, 
$\langle 1 1 1 \rangle$, and 7 cases for 
$\langle r_1 r_2 r_3 \rangle$, where $r_i$ are random numbers 
uniformly distributed in an interval of $[0,1]$.
The Velocity Verlet integration algorithm is utilized to model the 
collision cascade, which is performed for 6 ps, followed by an additional 
relaxation time of 4 ps. 
Electronic stopping, S$_e$ correction is included in our MD simulations due 
to the high PKA energy range considered in this work. The electronic stopping powers
\cite{ZBL} were obtained from SRIM-2013 \cite{SRIM-2013}
using the default values for material properties.

The MD simulations were done in the High Performance Computing Center of the
Max Planck Institute and the institutional cluster of the Stony Brook University 
by using the 
Large-scale Atomic/Molecular Massively Parallel Simulator (LAMMPS) 
\cite{PLIMPTON19951} with the Quantum mechanics and 
Interatomic Potential package (QUIP) \cite{quip} that is used as an interface to implement machine learned interatomic potentials
based on GAP \cite{PhysRevLett.104.136403}. 
We also perform MD simulations by using an embedded-atom method potential \cite{AcklandGJ, Salonen_2003}.
This MD potential is denoted as EAM in this work and has previously been applied to study 
the sputtering of single-crystalline Mo surfaces by Mo$_n$ ($n = 1, 2, 4$) projectiles 
in the total energy range of 0.125-4 keV.

In order to ensure accurate time integration in the high-energy collision dynamics.
We use an adaptive timestep that is implemented in LAMMPS with a maximum ($t_{max}$)  
of  4 fs. to assure that the maximum displacement of the Mo atoms per step is less 
than 0.01 \AA{}. In Tab. \ref{tab:tab1}, we report the numerical parameters for performing the 
MD simulations, and the time step used in the simulations as a function of 
the PKA.

\begin{table}[!b]
\caption{\label{tab:tab1} Size of the numerical boxes in 
nm as a function of the impact energy (PKA energy), which are used in the MD 
simulations. 
The lattice constant of the bcc Mo sample 
for GAP is $a=3.168$ \AA{} at 
$300$ K.}
\begin{indented}
\item[] \begin{tabular}{@{}lll}
\br
\textbf{PKA (keV)} & \textbf{Num. atoms} & \textbf{Box size (nm)} \\
\mr
0.5-2 &  25 392   & (7.24, 7.24, 7.55)  \\
5     & 55 800    & (9.51, 9.51, 9.82)  \\ 
10  & 104 044   & (11.72, 11.72, 12.04)  \\
\br
\end{tabular}
\end{indented}
\end{table}

\subsection{Identification of point defects and vacancies}
\label{sec:DV_method}

The damage in the Mo sample is analyzed by a software workflow for 
fingerprinting and visualizing defects in damaged crystal structures (FaVAD) 
\cite{2004.08184,favad}, where the local atomic environment of the $i$-th atom of 
the material sample 
is described by a descriptor vector (DV), $ \vec{\xi}^{\ i}$.
Here, the atom density around the $i$-th atom is expressed by a sum of a 
truncated Gaussian density functions with the difference vector $\vec r^{\ ij}$
between the atoms $i$ and $j$, entering the exponent \cite{PhysRevB.87.184115},

\begin{eqnarray}
\rho^{{} i}(\vec r) & = & \sum^{\textrm{neigh.}}_{j} \exp 
\left( -\frac{|\vec r-\vec r^{\ ij}|^2}{2 \sigma^2_{\textrm{atom}}} \right) 
f_{\textrm{cut}} \left( |\vec r^{\ ij}| \right)  \label{eq:Eq1} \\
     & = & \sum_{nlm}^{NLM} c^{(i)}_{nlm}g_n(r)Y_{lm}\left(\hat r\right),
\label{eq:Eq2}
\end{eqnarray}
which is then approximated in terms of spherical harmonic functions, $Y_{lm}(\hat r)$, 
and a set of basis functions in radial directions $g_n(r)$
as $c^{(i)}_{nlm} = \langle g_n Y_{lm} | \rho^i \rangle$ 
\cite{PhysRevB.87.184115,Jav_UvT}. 
The sum over the order $m$ of the squared modulus of the coefficients $c_{nlm}$
is invariant under rotations around the central atom \cite{PhysRevB.90.104108}.
It is given by
\begin{equation}
\vec{\xi}_{k}^{\ i} = \left\{ \sum_m
\left(c_{nlm}^i \right)^* c_{n'lm}^i \right\}_{\ n,n',l},
\label{eq:Eq3}
\end{equation}
where $c^{*}_{nlm}$ denotes the complex conjugate of $c_{nlm}$.
Here each component $k$ of the vector $\vec{\xi}$ corresponds to one of the index triplets $\{n,n',l\}$.
The normalized DV $\vec{q}^{\ i} = \vec{\xi}^{\ i}/|\vec{\xi}^{\ i}|$ is used throughout 
this work and calculated within the multi-body descriptor framework called 'Smooth Overlap 
of Atomic Positions' (SOAP), which implements Eqs. \ref{eq:Eq1} to \ref{eq:Eq3} 
with the Gaussian Approximation potential (GAP) \cite{PhysRevLett.104.136403}.

Once the DVs of all the atoms of the damaged material are computed, 
we calculate the distance between the two corresponding DVs, 
$d = d \left( \vec{q}^{\ i}, \vec{q}^{\ j} \right)$ to the atomic 
local environment of a defect free and thermalized material sample 
\cite{Jav_UvT,2004.08184} as follows
\begin{equation}
    d^M (T) =
    \sqrt{ \left( \vec{q}^{\ i} - \vec{v}\left(T\right) \right)^{\textrm{T}} 
\Sigma^{-1} (T) \left( \vec{q}^{\ i} - \vec{v}\left(T\right) \right)},
\label{eq:maha}
\end{equation} 
where $\vec{v}\left(T\right)=\frac{1}{N}\sum_{i=1}^{N}\vec{q}^{i}\left(T\right)$ is 
the mean DV of the defect free sample; and $\Sigma$ is the associated 
co-variance matrix of the DV components \cite{Maha,Jav_UvT}.
This calculation allows us to identify atoms outside of lattice positions that 
are then classified as point defects.
The identification of vacancies is done by defining a numerical sampling 
grid of $N=N_x\times N_y \times N_z$ points with $N_x,N_y,N_z$ being the number 
of equispaced points in the $x,y$, and $z$ directions, respectively \cite{Jav_UvT}. 
Followed by a computation of the nearest neighbor distance between the 
position of the damaged sample atoms and the sampling grid points.
Points where the distance to the nearest atom exceeds a given threshold describes 
the spatial volume of the identified vacancy \cite{2004.08184,favad}.

\section{Results and Discussion}
\label{results}

\begin{figure}[!b]
   \centering
   \includegraphics[width=0.5\textwidth]{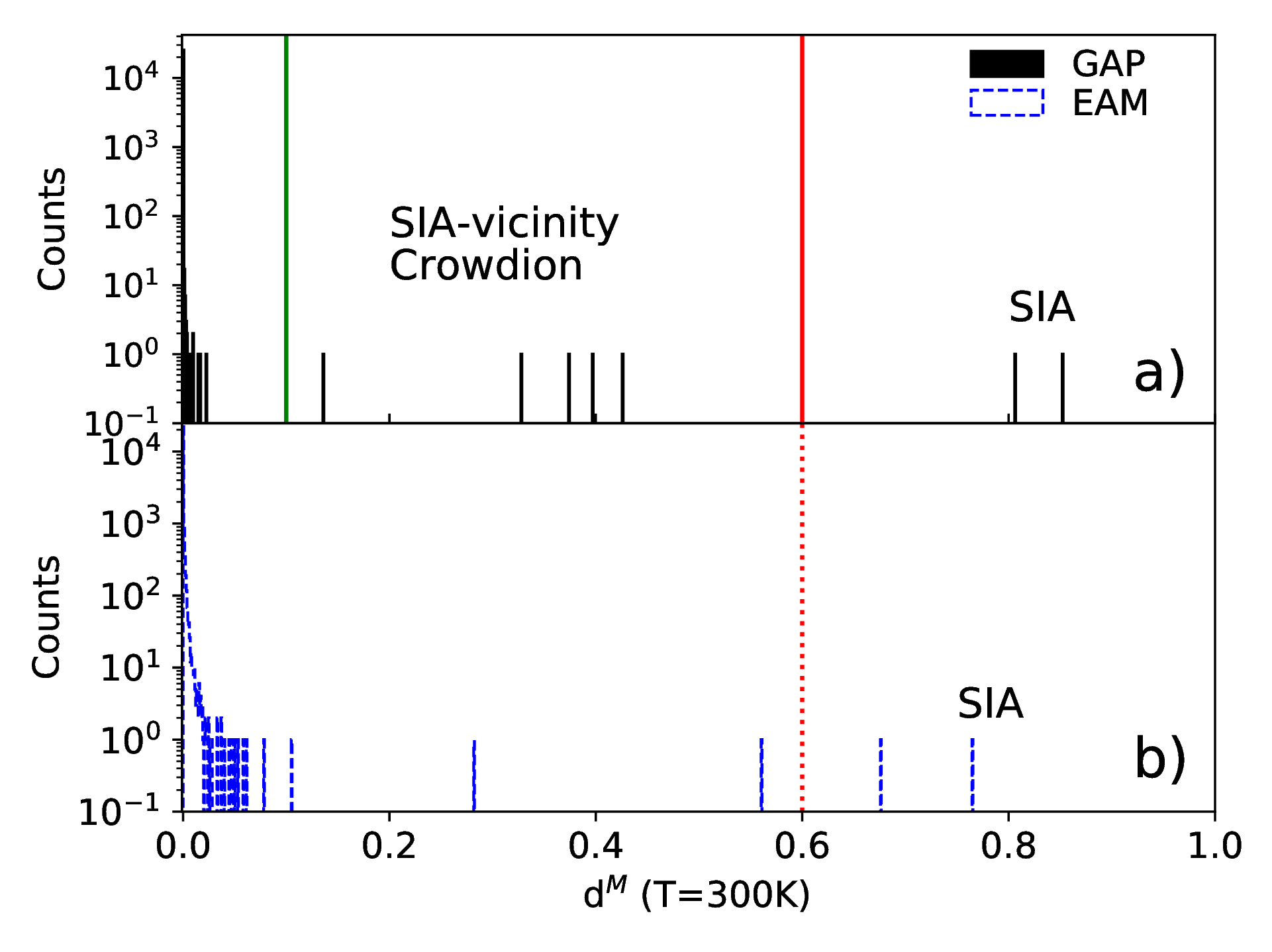}
   \caption{(Color on-line). Distance difference between the DVs for the damaged Mo sample 
   and those for the defect free and thermalized sample for the GAP in a) and EAM in b) 
   potentials. We apply FaVAD to a bombarded sample at a PKA energy of 500 eV on the 
   $\langle 001 \rangle$ velocity direction. 
   We notice that GAP potentials modeled the formation of 2 crowdions, while the 
   EAM potentials model the presence of only one crowdion defect after the collision cascade.}
   \label{fig:fig1}
\end{figure}

In order to analyze the damage in the Mo sample. 
We first perform a single MD simulation at 500 eV on the $\langle 001 \rangle$ velocity 
direction, with the GAP and EAM potentials for 10 ps of simulations time.
The final frame of the simulation contains the information of the point defects 
formed at the lowest PKA energy value of our study. 
Then, we analyze the damaged Mo sample by computing the DVs of all the Mo atoms 
with FaVAD, followed by the comparison to a defect free sample that is
thermalized to 300 K.

In Fig. \ref{fig:fig1} we present results for the distance difference, $d^M (300K)$,
between the DVs of the damaged Mo materials and the set of reference DVs obtained 
by using FaVAD.
The MD simulation performed with GAP presents two Mo atoms with the largest distance difference,
that allows us to set a threshold at 0.6 for further analyzes. 
A Mo atom with distance difference bigger than a value of 0.6 is quantified as 
a self-interstitial atom (SIA) for all the MD simulations.
We also identify Mo atoms in the vicinity of the SIA with a distance 
difference in the range of 0.1 to 0.6.
Although these atoms are not quantified as SIA, they provide information 
about the atomic arrangement of more complex point defects like a crowdion in this case, 
where four atoms share three lattice sites.
For the EAM case, two Mo atoms are also identified as potential SIA by FaVAD. 
However, the atomic geometry of this point defects is associated to a dumbbell defect 
where two atoms share a lattice site, observed in the Fig. \ref{fig:fig1}b) where only 
a couple of Mo atoms have a distance difference in the range of 0.2 to 0.6.
The GAP and EAM MD simulations report the same number of SIA, but 
the atomic geometry of the modeled defects is different. 
Besides that, the re-crystallization of the Mo sample is well modeled by the 
GAP potentials, where the majority of the Mo atoms have an atomic local environment
similar to the the defect free and thermalized Mo sample, which is noted in the 
upper panel of the same figure. 

\subsection{Crystal defects formation as a function of the simulation time}

The interatomic MD potentials need to be capable to model the formation of point defects
at different PKA energies assigned to the projectile. 
More complex defects can be found for high PKA energy values.
For this reason, we increase the features of FaVAD to compute the DVs of all the atoms 
of the sample at different time steps of MD simulations. 
Point defects can be identified by comparing their DVs to those for the defect-free sample 
thermalized to 300 K.
In order to show this new feature, we perform MD simulations at PKA energies of 10 keV and
1 keV on the $\langle 001 \rangle$ velocity direction. 
The information of the damage of the material is obtained as an output data from 
the MD simulation at a time step of $\Delta t_d =$ 0.1 ps for 1-3 ps, where the collision cascade 
mechanisms mainly happen \cite{BELAND2016136}; a $\Delta t_d =$ 0.5 ps for 3-6 ps; and a 
final $\Delta t_d =$ 1.0 ps for relaxing the damaged sample.

\begin{figure}[!t]
   \centering
   \includegraphics[width=0.48\textwidth]{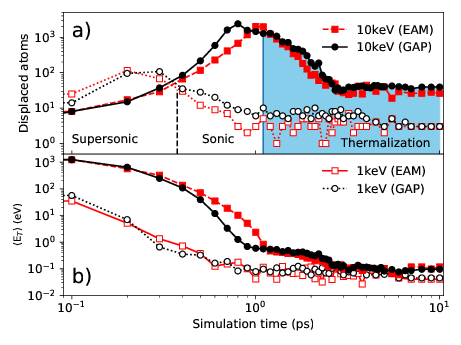} 
    \includegraphics[width=200pt, height=160pt]{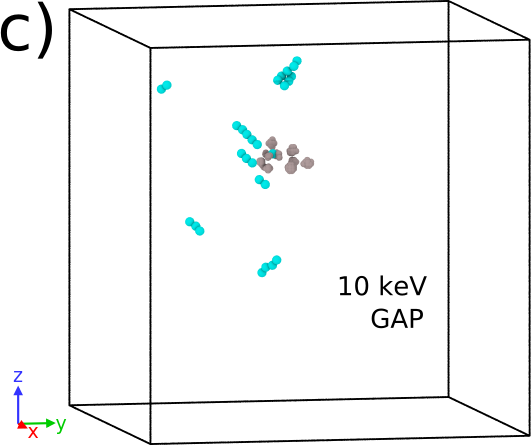}
   \caption{(Color on-line). Number of displaced atoms as a function of the simulation 
   time at 1 and 10 keV of PKA on the $\langle 001 \rangle$ velocity direction in a), by 
   using the GAP and EAM potentials. 
   At 10 keV, three phases are identified during collision dynamics associated with 
   the shockwave velocity where the destructive phase shows particles traveling at 
   supersonic velocities. Then the re-crystallization of the Mo sample is done during 
   the thermalization phase.
   The average of the kinetic energy of displaced Mo atoms are shown in b), 
   we notice a similitude between the two MD potentials. 
   Identified SIA and Mo atoms in the local vicinity (green spheres), as well as vacancies (Grey 
   spheres) are at the final step of the MD simulation with GAP are presented in c). 
   Noticing the formation of crowdion defects.}
   \label{fig:fig2}
\end{figure}

In Fig. \ref{fig:fig2}a), we present results for the quantification of point defects formed 
as a function of the simulation time at 1 (empty symbols) and 10 (solid symbols) keV of PKA 
with the GAP and EAM potentials.
FaVAD is applied to identify and quantify the point defects with a distance difference 
threshold of $d^M = 0.6$ for all cases, where is observed that Mo atoms have the highest 
probability to be considered as actual defects \cite{DOMINGUEZGUTIERREZ2020100724}. 
Although the profiles presented by the MD simulations at 10 keV are similar, 
the maximum number of displaced atoms is located at 0.8 and 1.1 ps for the GAP and EAM 
respectively. 
At the end of the MD simulations, a total of 42 displaced atoms are reported for the 
GAP modeling and 35 Mo atoms are identified as defects by utilizing the EAM potential.
The geometry of the identified point defects for the GAP MD simulations are shown in 
Fig. \ref{fig:fig2}c), where SIA and crowdions are presented by green spheres and vacancies are 
depicted as gray spheres.
The machine learning based MD potential produces a majority of crowdion defects rather than 
dumbbells, as will be discussed later.
At a PKA of 1 keV, the dynamics presented by the displacement atoms is notable different 
at 1-7 ps where the re-crystallization of the material sample is carried out. 
Being well modelled by the GAP framework (Fig. \ref{fig:fig2}a).
However, both MD potentials report the same number of three stable point defects at the end 
of the MD simulation.

In Fig. \ref{fig:fig2}a), three phases of a collision cascade are identified for cascades 
at a PKA energy of 10 keV by computing the average velocity, $\langle V(t) \rangle$, of 
the total identified displaced atoms, $N_D$, as a function of the simulation time as
\begin{equation}
    \langle V(t) \rangle = \frac{1}{N_D}  \sum_{i=1}^{N_D} \sqrt{v^i_x(t)^2 + v^i_y(t)^2+v^i_z(t)^2},
\end{equation}
where $v_x, v_y,$ and $v_z$ are the instantaneous velocities of the $i$-th 
displaced atom taken from the output data of the MD simulations.
By considering the longitudinal velocity of sound $\nu_\mathrm{Mo}$ in Mo that varies from 
5.4 to 6.25 km/s at a temperature of 300 K \cite{Dickinson}, we can define a supersonic 
collision cascade phase where $\langle V(t) \rangle > \nu_\mathrm{Mo}$.
Thus, at the beginning of the MD simulation (0.1-0.4 ps) the average velocity of the 
displaced atoms is 1.5 to 3 times bigger than $\nu_\mathrm{Mo}$, which is a similar phase range 
reported for iron samples \cite{CalderBacon} and Fe-Ni alloys \cite{BELAND2016136}.
In this supersonic phase, highly energetic atoms start colliding and transferring kinetic energy 
to their nearest neighbor atoms.  
At this lapse of time, atoms with supersonic velocities creates stable point defects in the sample, 
then some of them can be identified as Frenkel pairs, for example. 
More Mo atoms are displaced from their lattice position and the starting kinetic energy of the 
projectile is distributed among more Mo atoms, followed by loss of kinetic energy by 
the Mo atoms that were already displaced. 
This process lead the Mo sample to the Sonic phase in the time lapse of 0.4 to 1.1 ps. 
The sonic wave does not create stable point defects, 
leading to a liquid phase inside the Mo 
sample which is well modeled by the GAP potential due to the inclusion of liquid samples at different pressures 
in the training data \cite{2006.14365}.
For collisions at 10 keV, we noticed that the number of displaced atoms at the limit between
the supersonic and sonic phases is proportional to the total number of SIA at the end of 
the simulation time, as expected and reported in the literature for damage in materials
\cite{BELAND2016136}.
The last phase of the collision simulation is called the thermalization phase and
defined when the average velocity of the displaced atoms is $\langle V(t) \rangle < \nu_{Mo}$. 
In this phase the material sample is re-crystallized and cools down to its initial 
room temperature.

In Fig. \ref{fig:fig2}b) we report results for the average 
kinetic energy (KE) of displaced Mo atom as a function of the simulation time for PKA energy 
values of 1 and 10 keV with the GAP and EAM potentials.
The KE of the $i$-th displaced Mo atom is calculated as $E^i_K = \left( m/2 \right) 
\left[v^i_x(t)^2 + v^i_y(t)^2+v^i_z(t)^2 \right]$ with $m$ as the Mo mass. 
Followed by the computation of the average KE of all the displaced Mo atoms as $\langle E 
\rangle = \left( 1/N_D\right) \sum_i^{N_D} E^i_K$.
We notice that at the beginning of the MD simulation the projectile transfers its kinetic 
energy to its nearest neighbor atoms which starts the supersonic shock-wave. 
After 1.1 ps of simulation time, the average kinetic energy is almost constant, which 
represents the thermalization process of the sample for all the MD simulations at both PKA 
energy values.
However, EAM and GAP model the expansion of the sonic wave and the energy landscape of defects differently, which leads 
to the formation of different atomic geometry for the identified SIAs and Mo atoms in 
their vicinity at the end of the MD simulation. 
The formation of dumbbell SIA defects is more favorable in the MD simulations with 
the EAM potential.

\subsection{Classification and quantification of crystal defects 
as a function of the PKA energy}

\begin{figure}[!t]
   \centering
   \includegraphics[width=300pt,height=200pt]{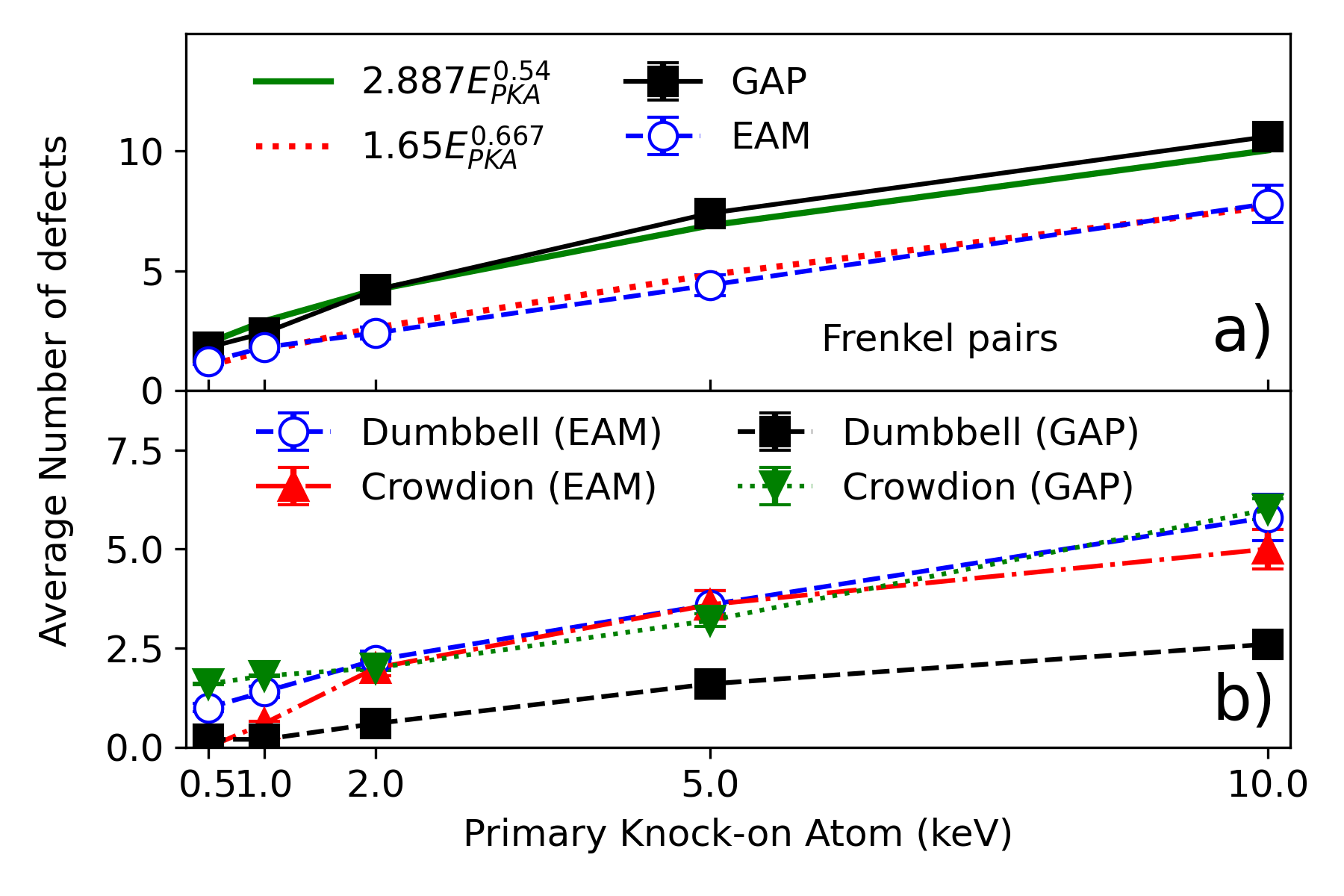}
   \caption{(Color on-line) Average number of point defects formed after collision cascade 
   as a function of the PKA energy. 
   Total number of Frenkel pairs in a), and crowdions and dumbbells in b).
   We compare results obtained by GAP with those for the EAM MD potentials. 
   A fitting curve is included to the total number of Frenkel pairs as: $a E_{PKA}^{b}$
   with $a = 2.887$ and $b = 0.54$ for the GAP and $a=1.65$ and $b=0.667$ for the EAM, 
   with a correlation factor of $0.99$ for both methods. Showing that number of 
   Frenkel pair increases with $E_{PKA}$.}
   \label{fig:fig3}
\end{figure}

\begin{figure}[!b]
   \centering
   \includegraphics[width=0.65\textwidth]{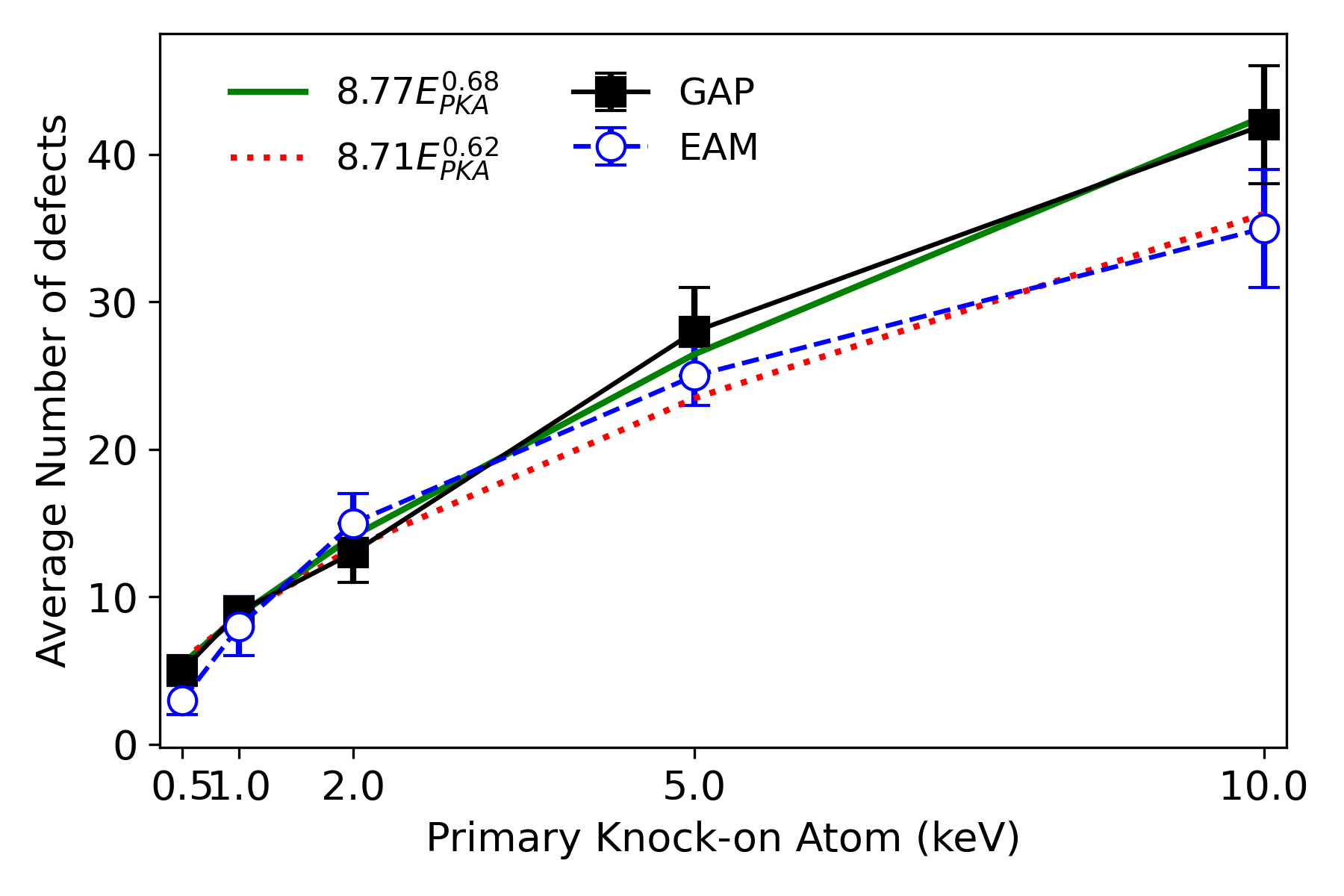}
   \caption{(Color on-line) Total number of defects calculated as: 
   Frenkel Pairs+ 3 $\times$ Crowdion + Dumbbell + SIA vicinity are presented
   as a function of the PKA.
   A fitting curve is included to the total number of defects as: $a E_{PKA}^{b}$;
   for GAP $a = 8.77$ and $b = 0.68$, while for EAM $a=8.71$ and $b=0.62$. A 
   correlation factor of $0.99$ is associated to the fitting curve for both methods. 
   }
   \label{fig:fig4}
\end{figure}

We perform MD simulations in a PKA energy range of 0.5 to 10 keV for several 
velocity directions with the GAP and EAM potentials.
The comparison of the modeling of the formation of point defects after collision 
cascade modeled by these potentials provide an insight of the advantages and limits 
of the GAP MD potentials over traditional ones (EAM potentials).
In Fig. \ref{fig:fig3} and \ref{fig:fig4} we present the average number of point
defects, $\langle PD (E_p)\rangle$, and its standard deviation, 
$\sigma(E_p)$ as a function of the PKA, $E_p$, which are calculated as:
\begin{eqnarray}
    \langle PD(E_p) \rangle &=& \frac{1}{N_T} \sum_{i=1}^{N_T} N_i(E_p|\langle r_1 r_2 r_3 
    \rangle) \nonumber \\
    \sigma(E_p) & =& \sqrt{\frac{1}{N_T-1} \sum_{i=1}^{N_T} \left(N_i(E_p|\langle r_x r_y r_z 
    \rangle - \langle PD(E_p) \rangle \right)^2}
\label{eq:average}
\end{eqnarray}
with $N_i(E_p|\langle r_1 r_2 r_3 \rangle)$ as the number of defects for a given 
velocity direction, $\langle r_1 r_2 r_3 \rangle$ , and $N_{T}$ as the total number 
of MD simulations performed. 
The defects are identified by FaVAD  as a function of the PKA energy. 

Fig. \ref{fig:fig3}a) shows the average number of Frenkel pairs as a function of the PKA
energy, modeled by the GAP and EAM potentials.
This value is related to the average number of single vacancies found in the 
damage Mo sample.
FaVAD has to identify the formation of a stable SIA and a vacancy to 
quantify this kind of defect. 
Usually Mo atoms with a $d^M (T) > 0.6$ are identified by FaVAD as SIA and 
quantified by using Eq. \ref{eq:average}, and Mo atoms in their vicinity 
indicate the formation of a crowdion or dumbbell defects.
The average number of Frenkel pairs can be fitted to a scaling law proposed by 
Setyawan et al. \cite{SETYAWAN2015329} that quantifies the number of 
point defects formed in damaged samples as: $a E_\mathrm{PKA}^{b}$, 
where $E_\mathrm{PKA}$ is the PKA energy value, and $a$ and $b$ are fitting parameters. 
In our case, a good approximation to our GAP data is found by applying the damped
least-square method with fitting parameters $a = 2.887$ and $b = 0.54$, 
where the associated correlation factor is $0.99$. 
We also fit the EAM results with $a=1.65$ and $b=0.667$ and a correlation factor of $0.99$.

\begin{table}[!t]
\caption{\label{tab:tab2} Average number of point defects and vacancies 
as a function of the PKA, which are identified by our DV based method. 
SIA are identified as W atoms with the largest $d^M(T)$ to be in an interstitial 
site, reported into parentheses. 
SIA vicinity are affected Mo atoms due to the presence of a SIA and thermal motion.
}
\begin{indented}
\item[ ]\begin{tabular}{@{}llllll}
\br
\multicolumn{6}{c}{GAP potential} \\
\mr
& \multicolumn{5}{c}{PKA (keV)}\\
\mr
\textbf{Defect} & \textbf{0.5} & \textbf{1} & \textbf{2} & \textbf{5} 
& \textbf{10}  \\
\mr
Frenkel Pairs   &  $2 \pm 1$ & $3 \pm 1$ & $4 \pm 1$  
&  $7 \pm 2$  &  $10 \pm 2$    \\
Crowdion   & $1 \pm 1$ & $2 \pm 1$ & $2 \pm 1$   & $3 \pm 2$   &  $6 \pm 2$     \\
Dumbbell        & 0 & 0 &   1 $\pm$ 1  & 2 $\pm$ 1   &  3 $\pm$ 2    \\
SIA Vicinity        & 0 & 0 & 2 $\pm$ 1 & 10 $\pm$ 1 & 11 $\pm$ 2 \\
\mr
Total           & $5 \pm 1$ & $9 \pm 1$ & $13 \pm 2$ & $28 \pm 3$  & $42 \pm 4$     \\
\br
\multicolumn{6}{c}{AT-EAM-FS potential} \\
\mr
 & \multicolumn{5}{c}{PKA (keV)}\\
 \mr
\textbf{Defect} & \textbf{0.5} & \textbf{1} & \textbf{2} & \textbf{5} 
& \textbf{10}  \\
\mr
Frenkel Pairs   & $1 \pm 1$ & $2 \pm 1$ &  $3 \pm 1$ 
& $4 \pm 1$  &   $8 \pm 2$    \\
Crowdion   & 0 & $1 \pm 1$ &  $2 \pm 1$  & $4 \pm 1$  & $5 \pm 2$     \\
Dumbbell         &  $1 \pm 1$ & $2 \pm 1$ &   $2 \pm 1$  &  $4 \pm 1$  & $6 \pm 2$      \\
SIA Vicinity     & $1 \pm 1$ & $1 \pm 1$ & 4 $\pm$ 1 & 5 $\pm$ 1 & $6 \pm 4$ \\
\mr
Total           & $3 \pm 1$ & $8 \pm 2$  &  $15 \pm 2$  &  $25 \pm 2$ & $35 \pm 4$   \\
\br
\end{tabular}
\end{indented}
\end{table}

In Fig. \ref{fig:fig3}b) we show results for the average number of dumbbells and 
crowdions as a function of the PKA energy. 
Here, the number of crowdions are quantified by identifying four Mo atoms sharing
three lattice positions, 
while dumbbells are detected when two Mo atoms share one lattice 
position by FaVAD \cite{DOMINGUEZGUTIERREZ2020100724}, with a $d^M (T) > 0.2$.
We notice that the formation of crowdions is more favorable for collision cascades 
simulated by the GAP potential, while dumbbells are formed for all the PKA energy 
values with the EAM potential. 
This is not a surprise, since the EAM potential incorrectly predicts $\langle 110 \rangle$ 
dumbbells to be lower in energy than $\langle 111 \rangle$ crowdions or dumbbells 
\cite{AcklandGJ}. 
The GAP describes the relative formation energies of all SIA defects in a good agreement 
with Density Functional Theory results \cite{2006.14365}, and correctly reproduces the 
$\langle 11\xi \rangle$ dumbbell as the most stable SIA \cite{PhysRevMaterials.3.043606}. 
The $\langle 11\xi \rangle$ dumbbell is a tilted $\langle 111 \rangle$ SIA that easily 
migrates along the $\langle 111 \rangle$ chain of atoms in a zigzag-like motion
\cite{PhysRevMaterials.3.043606}, fluctuating between dumbbell and crowdion 
configurations, both of which FaVaD detects and identifies. 

In Table \ref{tab:tab2}, we present the number of crystal defects as a 
function of the PKA energy for GAP and EAM MD simulations, as a reference.
Frenkel pairs (SIA + single vacancy) are counted for Mo atoms with 
maximum DV distance difference. 
Single vacancies are quantified and identified by the 
KD-tree algorithm included in FaVAD.
Identified crowdions and dumbbells are also tabulated.
The total number of defects is calculated as: Frenkel Pairs+ 3 $\times$ Crowdion 
+ Dumbbell + SIA vicinity and are shown in Fig. \ref{fig:fig4}. 
There is a fair agreement between GAP and EAM MD results at PKA energies lower than
5 keV. 
However, for high impact energy values, the discrepancy between the obtained 
results is due to the formation of crowdions which is more favorable for the GAP 
potentials (10 keV).
A fitting curve is included to the total number of defects as: $a E_{PKA}^{b}$ 
with $a = 8.77$ and $b = 0.68$ for GAP and $a=8.71$ and $b=0.62$ for EAM, 
resulting in a correlation factor of $0.99$ for both MD potentials. 

\subsection{Ion beam mixing}

\begin{figure}[!b]
   \centering
   \includegraphics[width=0.48\textwidth]{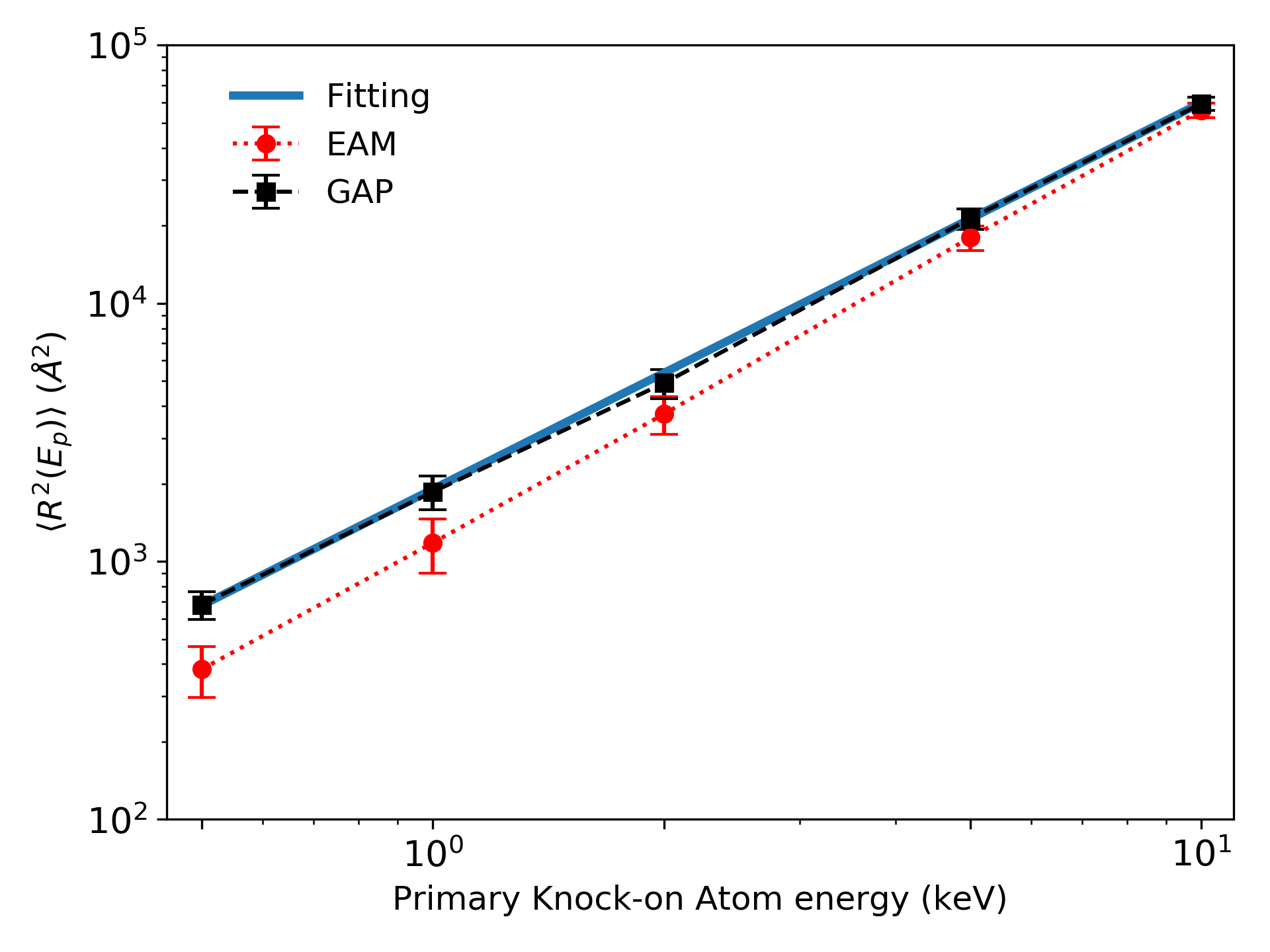}
   \caption{(Color on-line). The average of atom displacement R$^2$ as a function of cascade energy.
   Results calculated from MD simulations by GAP potential. 
   A function is fit to the data points as: $R^2(E_p) = a E_p^{3/2}$ with $a=1.865
   \times 10^3$ correlation factor of 0.99.}
   \label{fig:fig5}
\end{figure}

The damage in the Mo samples is also analyzed by computing the square of the total atom 
displacement, $R^2$, at the end of the collision cascade. 
This difference between the positions of the atoms at the end and the beginning 
of a cascade provides information about the formation of defects and 
the modeling of material re-crystallization. 
From the output data of the MD simulations, we can compute the atom displacement, $R^2 (E_p)$, as
 $\sum_i \left[ \vec r_i^f - \vec r_i^0 \right]^2$ \cite{doi:10.1063/1.366821}, where the index $i$ 
 runs for all the Mo  atom in the samples; and $\vec r^f$ and $\vec r^0$ are the positions of 
 each atom in the Mo sample at the final and first steps of the MD simulation, respectively. 
We verified that the center of mass of the cell is not displaced at the end of the MD simulations.
In Fig. \ref{fig:fig5} we report the average of the results, $\langle R^2 (E_p) \rangle$ as a function
of the PKA energy.
A fitting curve is approximated to our data points for the GAP MD simulations to 
possibly extrapolate values to higher impact energies \cite{BJORKAS20091830}. 
Since the PKA energy range considered in this work is $\leq$ 10 keV, the mixing is considered 
as a pure heat spike process. 
Letting to the GAP MD results be fitted to a simple power law as $ \langle R^2 (E_p) \rangle 
= a E_p^{\frac{3}{2}}$ with $a=1.865 \times 10^3$ as a fitting parameter with a correlation factor 
of 0.99. 
This analysis shows that more atoms are displaced as a function of the PKA, creating 
more vacancies and point defects in the damaged material as expected. 

We compute the mixing parameter $Q_{\textrm{sim}}$ from the MD simulation data 
by using the following expression \cite{PhysRevB.57.R13965}
\begin{eqnarray}
    Q_{\textrm{sim}}     & = \frac{R^2}{6n_0E_{D_n}}, \\
    \label{Eq:eq10}
\end{eqnarray}
where $n_0 = 2/a_0^3$ with $a_0 = 3.16$ \AA{} as the lattice constant of Mo,
$E_{D_n}$ is the deposited nuclear energy.
Obtained results are presented in Tab. \ref{tab:tab3}.
Allowing us to compare our MD calculations to experimental measurements of mixing 
efficiency defined as $Q_{\textrm{exp}}=Dt/ \Phi F_{D_n}$ following the atomistic 
definition of the diffusion coefficient \cite{BJORKAS20091830}. 
However, the quantitative comparison is not straightforward due to the irradiations are performed 
by using Kr ion with a mass of 83.79 amu and the computation of the experimental data is 
carried out by fitting Gaussian functions to the broadened marker layer shapes.

\begin{table}[!t]
\caption{\label{tab:tab3}
$Q_{\textrm{sim}}$ values as a function of the PKA.}
\begin{indented}
\item[] \begin{tabular}{@{}ll}
\br
PKA (keV) & Q$_{\textrm{sim}}$ (\AA{}$^5$/eV) \\
\mr
0.5  & 0.129   \\
1.0  & 0.172   \\
2.0  & 0.235   \\
5    & 0.340  \\ 
10   & 0.560  \\
\br
\end{tabular}
\end{indented}
\end{table}

\section{Concluding Remarks}
\label{conclusions}

In this paper, we performed classical and machine learning 
molecular dynamics simulations to emulate neutron bombardment 
on Molybdenum samples in an impact energy range of 0.5-10 keV, 
and a sample temperature of 300 K.
For this, we use a new machine learning interatomic potential
based on the Gaussian Approximation Potential (GAP) framework and results are 
compared to those obtained by using traditional Embedded Atom Method (EAM) potentials.
Formation of Frenkel pairs and more complex defects like crowdions and 
dumbbells are identified and quantified by using the recently developed 
software workflow for fingerprinting and visualizing defects in 
damaged crystal structures (FaVAD). 
Here, the local environment of each atom of the sample is represented by 
a descriptor vector. 
The difference between a pristine Mo sample and a damaged one is 
computed by considering thermal motion when the MD simulation is performed.

Super sonic, sonic, and thermalization phases are identified by 
a FaVAD features that compute the average kinetic energy of the identified 
defected Mo atoms.
The information of the liquid phase included in the fitting of the GAP potential leads to  
better modeling of the transition between super sonic and sonic phases, where 
complex defects start to form. 
The formation of crowdions is more favorable for the machine learning MD simulations.
EAM and GAP results from collision cascades simulations can be fit to a $\sim E_p^{b}$ 
scaling law, with $E_p$ as the PKA energy. 
Here $b=0.54$ for the GAP and $b=0.667$ for the EAM.

Machine learning interatomic potentials have been developed for more transitional metals like 
V, Nb, and Ta, where possible applications in damaged in material 
can be numerically modeled with high accuracy, which is part of our future work.

\ack
F.J.D.G gratefully acknowledges funding from A. von Humboldt 
Foundation and C. F. von Siemens Foundation for research fellowship.
Simulations were performed using the SeaWulf cluster 
at the Stony Brook University.
KN, FD and JB acknowledge that their part of this work has been carried out 
within the framework of the EUROfusion Consortium and has received funding 
from the Euratom research and training programme 2014-2018 under grant 
agreement No 633053. 
The views and opinions expressed herein do not necessarily reflect those of 
the European Commission.


\section*{References}
\bibliography{bibliography}

\providecommand{\newblock}{}
\begin{thebibliography}{10}
\expandafter\ifx\csname url\endcsname\relax
  \def\url#1{{\tt #1}}\fi
\expandafter\ifx\csname urlprefix\endcsname\relax\def\urlprefix{URL }\fi
\providecommand{\eprint}[2][]{\url{#2}}

\bibitem{doi:10.1063/1.2336465}
Rudakov D~L, Boedo J~A, Moyer R~A, Litnovsky A, Philipps V, Wienhold P, Allen
  S~L, Fenstermacher M~E, Groth M, Lasnier C~J, Boivin R~L, Brooks N~H, Leonard
  A~W, West W~P, Wong C~P~C, McLean A~G, Stangeby P~C, De~Temmerman G, Wampler
  W~R and Watkins J~G 2006 {\em Review of Scientific Instruments\/} {\bf 77}
  10F126

\bibitem{wirth_hu_kohnert_xu_2015}
Wirth B~D, Hu X, Kohnert A and Xu D 2015 {\em Journal of Materials Research\/}
  {\bf 30} 1440

\bibitem{Eren_2011}
Eren B, Marot L, Langer M, Steiner R, Wisse M, Mathys D and Meyer E 2011 {\em
  Nuclear Fusion\/} {\bf 51} 103025

\bibitem{LITNOVSKY20071395}
Litnovsky A, Wienhold P, Philipps V, Sergienko G {\em et~al.\/} 2007 {\em
  Journal of Nuclear Materials\/} {\bf 363-365} 1395 -- 1402

\bibitem{Nor17b}
Nordlund K, Zinkle S~J, Sand A~E, Granberg F, Averback R~S, Stoller R, Suzudo
  T, Malerba L, Banhart F, Weber W~J, Willaime F, Dudarev S and Simeone D 2018
  {\em Nature communications\/} {\bf 9} 1084

\bibitem{Nor18}
Nordlund K, Zinkle S~J, Sand A~E, Granberg F, Averback R~S, Stoller R, Suzudo
  T, Malerba L, Banhart F, Weber W~J, Willaime F, Dudarev S and Simeone D 2018
  {\em J. Nucl. Mater.\/} {\bf 512} 450--479

\bibitem{BOLT200243}
Bolt H, Barabash V, Federici G, Linke J, Loarte A, Roth J and Sato K 2002 {\em
  Journal of Nuclear Materials\/} {\bf 307-311} 43 -- 52

\bibitem{PhysRevB.29.6443}
Daw M~S and Baskes M~I 1984 {\em Phys. Rev. B\/} {\bf 29}(12) 6443--6453

\bibitem{AcklandGJ}
Ackland G~J and Thetford R 1987 {\em Philosophical Magazine A\/} {\bf 56}
  15--30

\bibitem{Salonen_2003}
Salonen E, rvi T~J, Nordlund K and Keinonen J 2003 {\em Journal of Physics:
  Condensed Matter\/} {\bf 15} 5845--5855
  \urlprefix\url{https://doi.org/10.1088%2F0953-8984%2F15%2F34%2F314}

\bibitem{DOMINGUEZGUTIERREZ2020100724}
Domínguez-Gutiérrez F, Byggmästar J, Nordlund K, Djurabekova F and von
  Toussaint U 2020 {\em Nuclear Materials and Energy\/} {\bf 22} 100724

\bibitem{Jesper_GAP}
Byggm{\"a}star J, Hamedani A, Nordlund K and Djurabekova F 2019 {\em Phys. Rev.
  B\/} {\bf 100} 144105

\bibitem{PhysRevLett.104.136403}
Bart\'ok A~P, Payne M~C, Kondor R and Cs\'anyi G 2010 {\em Phys. Rev. Lett.\/}
  {\bf 104}(13) 136403

\bibitem{doi:10.1080/21663831.2020.1771451}
Hamedani A, Byggmästar J, Djurabekova F, Alahyarizadeh G, Ghaderi R, Minuchehr
  A and Nordlund K 2020 {\em Materials Research Letters\/} {\bf 8} 364--372

\bibitem{2006.14365}
Byggmästar J, Nordlund K and Djurabekova F 2020 Gaussian approximation
  potentials for body-centered cubic transition metals (\textit{Preprint}
  \eprint{arXiv:2006.14365})

\bibitem{2004.08184}
von Toussaint U, Dominguez-Gutierrez F~J, Compostella M and Rampp M 2020 Favad:
  A software workflow for characterisation and visualizing of defects in
  crystalline structures (\textit{Preprint} \eprint{2004.08184})

\bibitem{favad}
 2020 \url{https://gitlab.mpcdf.mpg.de/NMPP/favad/-/tree/master/}

\bibitem{Jav_UvT}
Dom\'{i}nguez-Guti\'{e}rrez F and von Toussaint U 2019 {\em Journal of Nuclear
  Materials\/} {\bf 528} 151833

\bibitem{PhysRevB.87.184115}
Bart\'ok A~P, Kondor R and Cs\'anyi G 2013 {\em Phys. Rev. B\/} {\bf 87} 184115

\bibitem{PhysRevB.85.214121}
Park H, Fellinger M~R, Lenosky T~J, Tipton W~W, Trinkle D~R, Rudin S~P,
  Woodward C, Wilkins J~W and Hennig R~G 2012 {\em Phys. Rev. B\/} {\bf 85}(21)
  214121

\bibitem{WBPearson}
Pearson W~B 1967 {\em Handbook of Lattice Spacing and Structures of Metals\/}
  (Pergamon, Oxford)

\bibitem{DOMINGUEZGUTIERREZ201756}
Dom\'inguez-Guti\'errez F and Krsti\'c P 2017 {\em Journal of Nuclear
  Materials\/} {\bf 492} 56 -- 61

\bibitem{ZBL}
Ziegler J~F, Biersack J~P and Littmark U 1985 {\em {The Stopping and Range of
  Ions in Matter}\/} ({New York}: Pergamon)

\bibitem{SRIM-2013}
Ziegler J~F {SRIM-2013 software package, available online at
  \textit{http://www.srim.org}.}

\bibitem{PLIMPTON19951}
Plimpton S 1995 {\em Journal of Computational Physics\/} {\bf 117} 1 -- 19

\bibitem{quip}
 2018 \url{http://libatoms.github.io/QUIP/}

\bibitem{PhysRevB.90.104108}
Szlachta W~J, Bart\'ok A~P and Cs\'anyi G 2014 {\em Phys. Rev. B\/} {\bf 90}
  104108

\bibitem{Maha}
Mahalanobis P 1930 {\em J. and Proc. Asiat. Soc. of Bengal\/} {\bf 26} 541

\bibitem{BELAND2016136}
Béland L~K, Osetsky Y~N and Stoller R~E 2016 {\em Acta Materialia\/} {\bf 116}
  136 -- 142

\bibitem{Dickinson}
Dickinson J~M and Armstrong P~E 1967 {\em Journal of Applied Physics\/} {\bf
  38} 602--606

\bibitem{CalderBacon}
Calder A, Bacon D, Barashev A and Osetsky Y 2010 {\em Philosophical Magazine\/}
  {\bf 90} 863--884

\bibitem{SETYAWAN2015329}
Setyawan W, Nandipati G, Roche K~J, Heinisch H~L, Wirth B~D and Kurtz R~J 2015
  {\em Journal of Nuclear Materials\/} {\bf 462} 329 -- 337

\bibitem{PhysRevMaterials.3.043606}
Ma P~W and Dudarev S~L 2019 {\em Phys. Rev. Materials\/} {\bf 3} 043606

\bibitem{doi:10.1063/1.366821}
Nordlund K, Ghaly M and Averback R~S 1998 {\em Journal of Applied Physics\/}
  {\bf 83} 1238--1246

\bibitem{BJORKAS20091830}
Björkas C and Nordlund K 2009 {\em Nuclear Instruments and Methods in Physics
  Research Section B: Beam Interactions with Materials and Atoms\/} {\bf 267}
  1830 -- 1836

\bibitem{PhysRevB.57.R13965}
Nordlund K, Wei L, Zhong Y and Averback R~S 1998 {\em Phys. Rev. B\/} {\bf
  57}(22) R13965--R13968

\end{thebibliography}
\bibliographystyle{iopart-num}
\end{document}